# Smart Cities: Striking a Balance Between Urban Resilience and Civil Liberties

Sangchul Park (School of Law, Seoul National University, Seoul Korea)

Cities are becoming smarter and more resilient by integrating urban infrastructure with information technology. However, concerns grow that smart cities might reverse progress on civil liberties when sensing, profiling, and predicting citizen activities; undermining citizen autonomy in connectivity, mobility, and energy consumption; and deprivatizing digital infrastructure. In response, cities need to deploy technical breakthroughs, such as privacy-enhancing technologies, cohort modelling, and fair and explainable machine learning. However, as throwing technologies at cities cannot always address civil liberty concerns, cities must ensure transparency and foster citizen participation to win public trust about the way resilience and liberties are balanced.

**Introduction**

Most of our future generations will live in smart cities. The world is urbanizing rapidly; the rate of urbanization is expected to grow, between 2018 and 2050, from 79% to 87% in developed countries and from 51% to 66% in developing countries (UN 2019, p. 11). Cities, which cover only 2% of the earth's surface, reportedly account for approximately 70% of carbon emissions (UN 2020, pp. 18–19), 2/3 of energy consumption (*Id.*), and 70% of waste generation (Zaman and Lehmann 2013). Cities are also constantly challenged by societal problems such as crime (Glaeser and Sacerdote 1999), traffic problems (Buchanan 1963), public health risks (Corburn 2004), and social divides (OECD 2018). To increase their resilience against these problems, cities are transitioning to smart cities by integrating urban infrastructure with information and sustainable technologies (Naphade et al. 2011). Smart city initiatives are burgeoning worldwide, serving as cradles for innovation and test beds for our descendants' habitats (Figure 1).

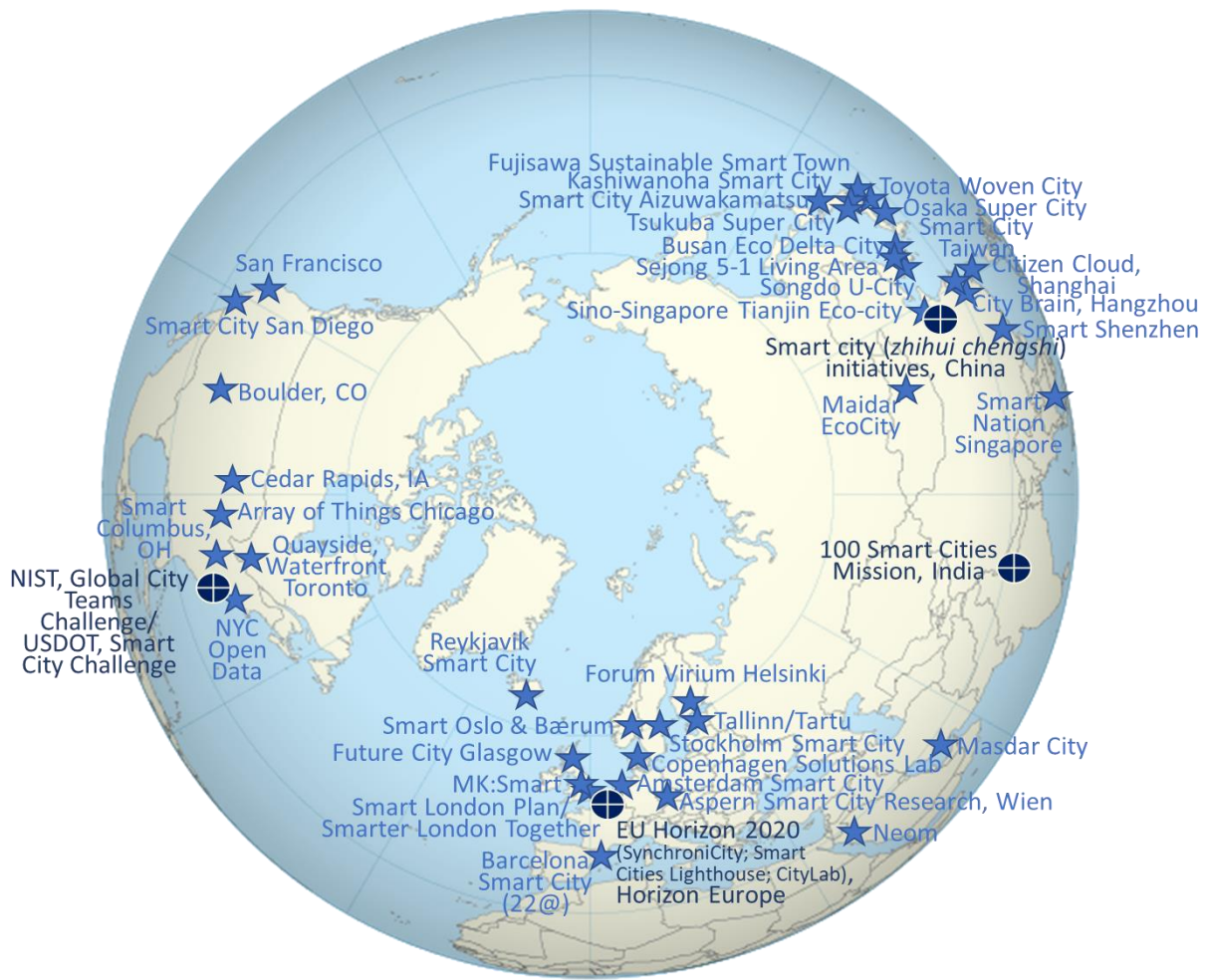

Figure 1. Overview of smart cities. Source: *Author* (cities) and *David A. Eccles* (background hemisphere map, distributed under a CC0 1.0 license.)

However, transitions to smart cities are subject not only to technical and fiscal constraints but legal and political challenges (Naphade et al. 2011). Smart cities sense, profile, and predict citizen activities; control connectivity and mobility; and reorganize energy, land, and digital resources, for enhanced urban planning including data-driven policing, intelligent transportation systems (ITSs), environmental monitoring, smart energy and waste management, education, and healthcare (Sookhak et al. 2019). During the COVID-19 pandemic, digital contact tracing and other emergency schemes offered a preview of legal and political challenges future cities must confront, not to mention South Korea (Korea)'s diversion of an existing smart city hub to a centralized tracing system

(Park, Choi, and Ko 2021) and Singapore's electronic monitoring (EM) of isolated or quarantined people as a component of its Smart Nation initiative (Smart Nation Singapore 2022). In this context, this article reviews ongoing debates over potential friction between smart city functions purporting to increase urban resilience and existing legal systems designed to support civil liberties.

**Sensing and Profiling: Privacy and Mass Surveillance Concerns**

A smart city runs 24/7 sensors including closed-circuit television (CCTV), IP cameras, internet-of-things (IoT) devices, and smart meters. They serve as the smart city's sensory organs, directing the city to urban actions for resilience. They also help fit the city's platform to the real world it purports to simulate. However, a smart city's potential to profile citizen activities through the sensed data has prompted fears about a citizen's "right to be let alone" (Warren and Brandeis 1890) being insidiously compromised by a "digital panopticon" or mass surveillance (Sadowski and Pasquale 2015), as portrayed in George Orwell's Nineteen Eighty-Four (1949).

*Consent Mechanisms*

It was believed that *consent* could address privacy concerns by enabling people to exercise control over the processing of their data. Under the EU's General Data Protection Regulation (GDPR), consent is now only one of six legal bases for data processing (art. 6(1)(a)) but still has cardinal practical importance due to the indefinite nature of other legal bases (such as "for the purposes of the legitimate interests pursued" (art. 6(1)(f))). Due to the influence the EU framework has exerted worldwide (except for the U.S. privacy law, which is relatively more sector-specific and harm-based), the consent principle is central to data protection laws in various jurisdictions. Pursuant to the legal framework, smart cities have tried to implement the following consent mechanisms.

- *Opt-in:* Opt-in or prior explicit consent is the most typical mechanism for compliance with data protection law; silence, pre-ticked boxes, and inactivity do not constitute consent under the GDPR (Recital 32). Smart cities built from the ground up can relatively easily obtain prior consent from each new resident. In 2020, Busan Eco Delta City included in its housing application a consent form for processing personal data for living labs (Busan EDC 2020). In 2021, Toyota-led Woven City also collected prior consent from every new resident (Chiba, Inoue and Miura 2021). However, those transitioning to smart cities face greater difficulties in obtaining consent. They often embed opt-in mechanisms in mobile apps developed to provide urban services, as illustrated by Aizuwakamatsu's disaster alert app (Chandran 2021). Regarding sensors installed in a public space, the "ubiquity of data collection and the practical obstacles to providing information without a user interface" (FTC 2015) makes it a daunting task to obtain consent from each passerby. This had cities consider implied consent schemes, as noted below.

- *Signage-based, or other interface-based consent*: This approach constructs implied consent from a citizen's behavior, such as entry into a monitored place or use of certain services after being afforded a chance to review signage, QR codes, and other interfaces (FTC 2015). Many data protection laws require not consent but disclosure from sensors installed for public duties. For example, if personal data are processed "for the performance of a task carried out in the public interest or in the exercise of official authority vested," the GDPR does not require consent (art. 6(1)(e)) but information provision (art. 12). Korea's Personal Information Protection Act (PIPA) (*Gaein-jeongbo-boho-beop*) exempts, from the consent requirement, cameras installed in the public space for certain public duties, on condition that signage is posted to provide privacy notices (arts. 25 and 58(2)). For the Sidewalk Toronto project, Google's Sidewalk Labs designed intuitive icons for signage (Sidewalk Labs 2019b) (Figure 2).

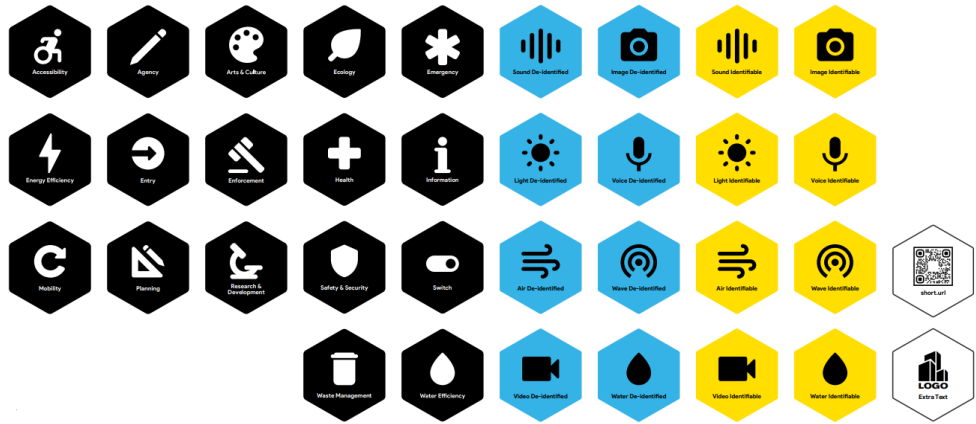

Figure 1. Icon designs developed by Sidewalk Labs. Source: *The Digital Transparency in the Public Realm project contributors*, distributed under a CC-BY 4.0 license

- *Opt-out*: Opt-out is a method by which citizens withdraw implied consent. Starting from September 2023, Korea's PIPA applies the opt-out scheme to photographing by mobile cameras (art. 25-2(1)(ii)). To place human samples into biobanks, Finland's Biobank Act (*Biopankkilaki*) (sec. 13) and France's Public Health Code 2022 (*Code de la santé publique 2022*) (art. L1131-1(1)) apply opt-out schemes, respectively, to the transfer of old samples to biobanks and the reuse of samples already collected (Kaye et al. 2016).

That said, the consent principle has evident limitations. Implied consent is, at best, a legal fiction which does not really address privacy risks. Explicit consent, as a condition of residence in or entry to a smart city or access to its infrastructure, might not constitute a citizen's voluntary consent based on a feeling of willingness. Mandated disclosures, such as privacy notices, can undermine consumers' protections and self-guard (Ben-Shahar and Schneider 2011). The pretext of consent easily exonerates cities from liability for privacy harms, resulting in a suboptimal level of precaution and self-regulation (Ben-Shahar 2019).

*Blanket Ban*

The scepticism about consent has had legislators consider "no-go" zones or a blanket ban on certain types of sensing that pose higher surveillance risks.

- *Real-time biometric identification*: Recent progress in computer vision and data availability vastly increased a city's capability to recognize biometric markers, including face, iris, retina, and behavior patterns for identifying pedestrians. The EU is moving toward restricting biometric identification. Under the GDPR, the "biometric data for the purpose of uniquely identifying a natural person" falls under "special categories of personal data" or sensitive personal data (art. 9(1)), the processing of which requires explicit consent from the data subject (art. 9(2)(a)) and is not exempt from purpose limitation (art. 6(4)(c)). As it is practically impossible for a city to obtain explicit consent from each passerby, the real-time biometric identification of citizens is almost banned. The European Commission (EC)'s proposal for the Artificial Intelligence Act (draft AI Regulation) explicitly bans the use of artificial intelligence systems for real-time remote biometric identification in publicly accessible spaces for the purpose of law enforcement (art. 5(1)(d)). From 2019 to 2021, several U.S. states and municipalities, including Virginia, Boston, and San Francisco, passed legislation curbing government use of facial recognition (Lee and Chin 2022). However, amid a surge in crime, many have undone or are undoing this; Vermont, the last state retaining a blanket ban over law enforcement use of facial recognition, stepped back by allowing it for investigating child sex trafficking (Reuters 2022).

- *Location tracking:* Smartphones, wearable devices, smart cards, and connected vehicles are ubiquitously tracking people's locations in real time. In societies based on liberal ideas, the public sector's massive tracking of citizens through access to their mobile devices would generally be unconstitutional. However, the exceptional permissibility of contact tracing or EM for public health or law enforcement remains debatable. At the outset of the pandemic, the European Data Protection Board (EDPB)'s Guidelines 04/2020 required contact tracing apps to use anonymized location data and thus conduct proximity tracing rather than track individual movements (EDPB 2020). Based on the guideline, Norway's data protection authority had the public health authority stop tracking GPS through a COVID-19 app (Datatilsynet 2020), and most European countries developed and released decentralized and anonymized proximity tracing apps. Several countries including Singapore and Korea ran EM apps to enforce isolation or quarantine orders (Park, Choi and Ko 2021; Smart Nation Singapore 2022). EM for law enforcement is justified as "community control" (State of Florida Office of the Auditor General 1993) with its legitimacy sought from cost-effectiveness and beneficial effects on recidivism (Henneguelle, Monnery and Kensey 2016) but at once denounced as "e-carceration" which facilitates social marginalization (Arnett 2019) and turns a smart city into a carceral city.

*Privacy-Enhancing Technologies*

All these circumstances have had cities divert attention from consent to *privacy-enhancing technologies*. "Data protection by design and default" under the GDPR (art. 25), previously known as "privacy by design" (Cavoukian 2011), requires smart cities to take technical measures, such as pseudonymization, to implement data protection principles such as data minimization (art. 25(1)). Different technologies have been deployed to defend data against reidentification attacks or "doxing."

*De-identifying data versus publishing aggregate information*

The first approach is *to preprocess data for de-identification* before feeding them into models.

- *Anonymization and pseudonymization*: These are the most viable measures for cities to process personal data without consent under the GDPR (art. 25(1)) and laws influenced by it. In the U.S., New York City (NYC)'s IoT guidelines stipulate that data should by default be anonymized before being shared (NYC 2017), and San Francisco's open data release toolkit recommends pseudonymization for most datasets (City and County of San Francisco 2016). Identification occurs when an identifier *singles out* a person, or a set of quasi-identifiers causes reidentification through *linkage* with another database or *inference* from auxiliary information (Article 29 Data Protection Working Party 2014). Hence, anonymization and pseudonymization typically delete or mask identifiers and suppress or generalize parts of quasi-identifiers (Samarati and Sweeney 1998) but differ in handling identifiers (Figure 3). As applied to computer vision, they often take the form of *obfuscation* such as pixelization and blurring, which is vulnerable to reidentification (McPherson, Shokri and Shmatikov 2016). As an alternative, *semantic segmentation* segments an image by categorizing features, such as pedestrians, vehicles, and traffic signs (Garcia-Garcia et al. 2017). The crux of anonymity is difficulty in predefining the degree of suppressing or generalizing quasi-identifiers (Samarati and Sweeney 1998).

An equivalence class whose elements have a set of quasi-identifiers in common satisfies $k$-anonymity if it has at least $k$ records (*Id.*), $\ell$-diversity if it has at least $\ell$ well-represented values for a sensitive attribute (Machanavajjhala et al. 2006), and $t$-closeness if its distribution of a sensitive attribute is no more distant than $t$ from that in the whole table (Li, Li and Venkatasubramanian 2007). Larger $k$ and $\ell$ and smaller $t$ values cause greater data utility loss, while not yet providing perfect privacy guarantees (Dwork and Roth 2013). The inability of laws to predefine $k$, $\ell$, and $t$ applicable to different cases creates legal uncertainty.

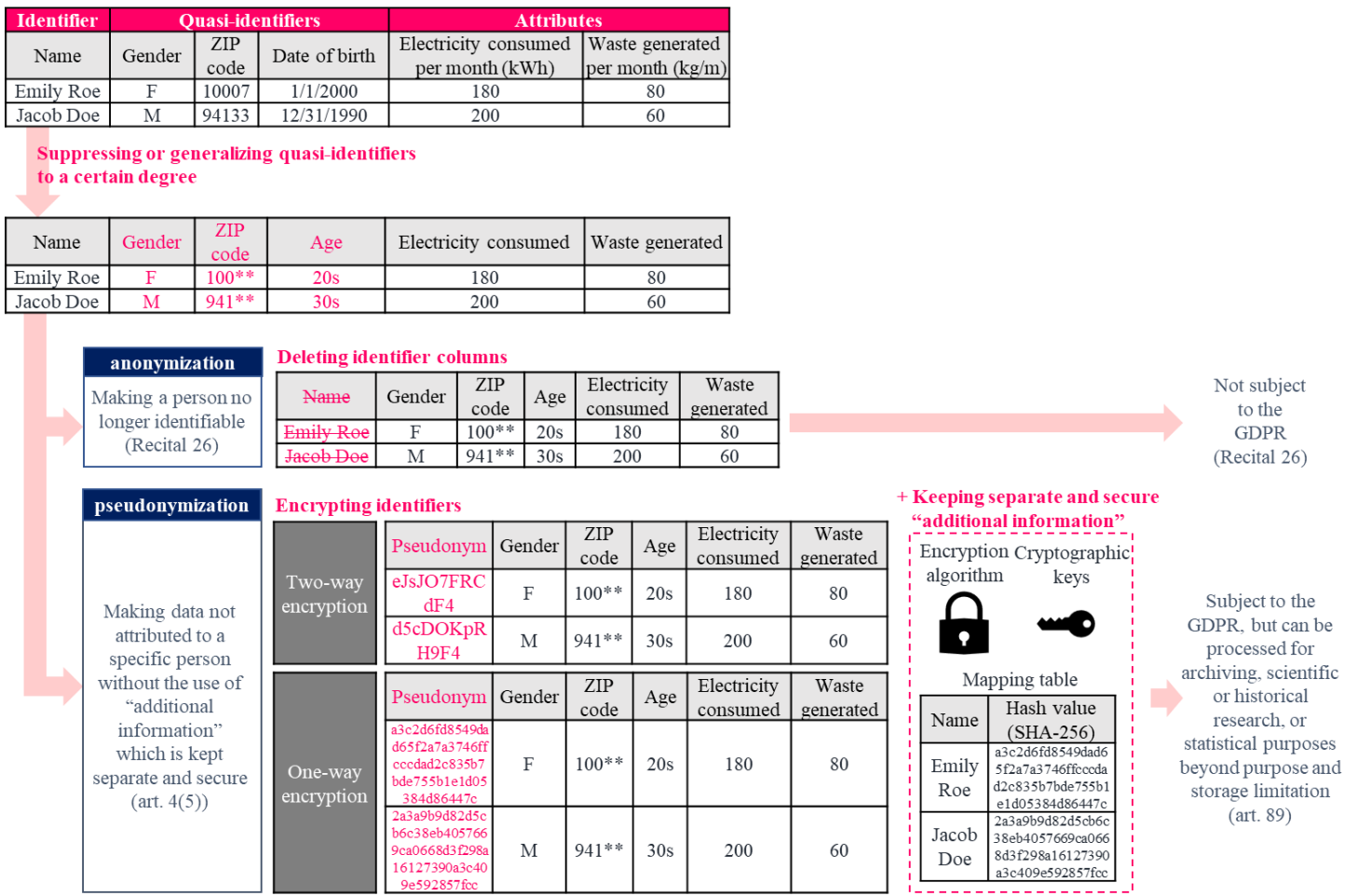

Figure 3. GDPR-compliant anonymization and pseudonymization. Source: *Author*

- *Secure computation*: *Homomorphic encryption (HE)* is a cryptosystem that enables a third party to compute on encrypted data without decrypting it (Gentry 2010). If a fully homomorphic encryption can be practically used through further studies on reducing its computational overhead, it could be a game changer that guarantees privacy without data utility loss (Li, Luo and Liu 2010). HE has been applied to smart grids (*Id.*) and a COVID-19 app (An et al. 2021) and can accelerate smart cities' open data initiatives by realizing the secure outsourcing of computation to citizens (Yang et al. 2019). *Multi-party computation* allows multiple parties to jointly compute a function by feeding a set of inputs hidden from other parties (Yao 1986).

- *Synthetic data (SD)*: SD is data artificially generated to yield statistical inferences close to those from the original dataset (Raghunathan 2021). SD is a versatile technology that can be used for oversampling, emulating environments, and transfer learning, but can also provide privacy guarantees without seriously impairing the validity of statistical inferences (*Id.*). The U.S. Census Bureau applied SD to employee commuting pattern data making them publicly available on its web application entitled OnTheMap (Machanavajjhala et al. 2008).

The second approach is to *publish aggregate information* while keeping individual data intact and non-inferable.

- *Differential privacy (DP)*: DP injects random noises into or otherwise randomly post-processes responses to queries so that the presence or absence of a single person's data does not affect the responses significantly enough to have the person inferable from them (Dwork 2008). To formalize, a randomized algorithm $\mathcal{M}: \mathcal{D} \to \mathcal{R}$ gives $\varepsilon$-DP, if $Pr[\mathcal{M}(d_1) \in S] \leq$

$e^\varepsilon Pr\;[\mathcal{M}(d_2) \in S]$ holds for any inputs $d_1, d_2 \in D$ differing on at most one element and for the set of possible outputs $S \subseteq \mathcal{R}$ (*Id.*). DP provides privacy guarantees against reconstruction attacks, while offering a formally defined metric ε (privacy budget) representing privacy loss at a marginal change in data (*Id.*). DP has been frequently used for urban mobility tracking (for highway traffic statistics by Google (Google 2015) and for data collected from the Opal smart ticketing system in Sidney (Asghar, Tyler and Kaafar 2017)) and smart metering (Eibl and Engel 2017; Hassan et al. 2019).

- *Federated learning (FL)*: FL is a machine learning technique that trains a global model by pooling parameters or hyperparameters that each local node (such as a smart device) learns from local data without directly accessing the data (Konecny et al. 2016). FL can be combined with DP to prevent parameters from singling out a person (Geyer, Klein, and Nabi 2017). FL has been extensively applied to essential components of smart city hubs such as the IoT, ITSs, and financial, medical, and communication services (Zheng et al. 2022).

*Decentralization versus centralization*

The degree of decentralization also has profound implications for privacy and data utility.

- *Decentralized model*: In a pure peer-to-peer (P2P) environment, a smart city's role is limited to developing and licensing apps; each smart device or personal computer held by citizens collects, processes, and exchanges personal data. The model confers stronger privacy protection, while data utility can be limited. Decentralized FL learns a model in a dynamic P2P environment without a server (Figure 4) (Roy et al. 2019), and the MPC typically works on decentralized nodes (Yao 1986). Public blockchain networks should be viewed as a centralized model comprising multiple nodes rather than a decentralized model, in respect to data that are broadcast and immutably recorded on chain (Eberhardt and Tai 2017).

- *Hybrid model*: From the perspective of system architecture, "local" DP (such as RAPPOR (Erlingsson, Pihur and Korolova 2014), succinct histogram (Bassily and Smith 2015), and local hashing (Wang et al. 2017)) and "centralized" FL (Hassan et al. 2019) can be classified as a hybrid model, in that original datasets are held locally, but perturbated query responses (in DP) or parameters (in FL) are pooled at a central system for more useful analysis (Figure 4). Edge computing sometimes refers to an architecture where data are de-identified near IoT sensors upon sensing or near local nodes storing the original dataset (Shi et al. 2016). Chicago, for its Array of Things project, analyzed the sensed images within sensor nodes through software-defined sensing and deleted them (Catlett et al. 2020).

- *Centralized model*: This model pursues greater data utility, running higher surveillance risks. An adjustment to this model includes outsourcing data pooling and de-identification to a trusted third party. Under a global DP model, a third-party curator aggregates data and adds noises to outputs of queries (Figure 4) (Dwork 2008).

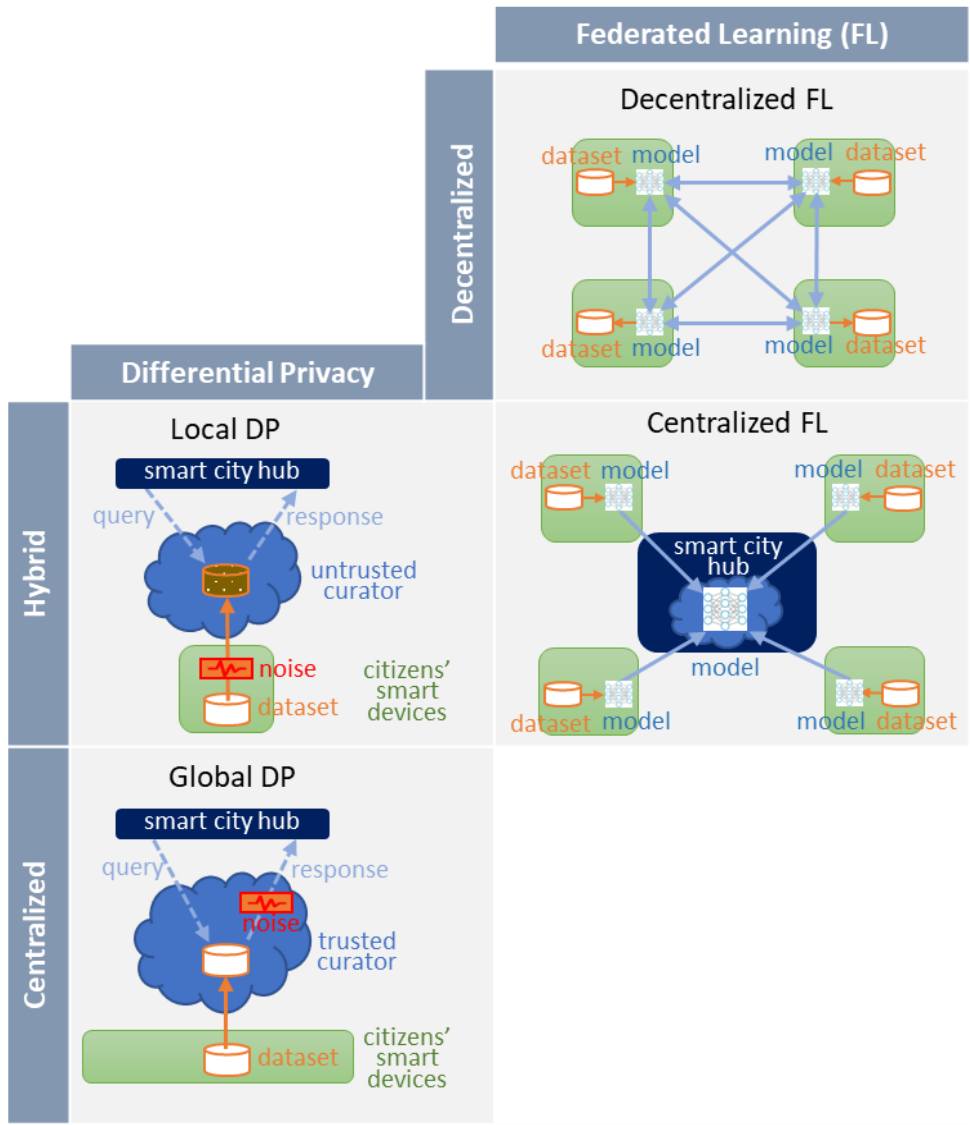

Figure 4. DP, FL, centralized model, and decentralized model. Source: *Author*

A key issue involving the COVID-19 digital contact tracing was balancing privacy with public health along the spectrums of decentralized and centralized models.

- *Decentralized proximity tracing apps*: Many Western countries developed and released proximity tracing apps backed by Google/Apple Exposure Notification API. Pursuant to the EDPB's Guidelines 04/2020, they implemented encounter logging by sending ephemeral IDs via Bluetooth Low Energy to nearby devices, and infection reporting by allowing confirmed cases to voluntarily broadcast their IDs to close contacts (EDPB 2020). Examples include Switzerland ("SwissCovid"), Italy ("Immuni"), Germany ("Corona-Warn-App"), England and Wales ("NHS COVID-19"), Japan ("COCOA"), and several U.S. states. Despite their privacy-enhancing nature, their efficacy against the pandemic has been dubious, due to their failure to reach a penetration rate sufficient to properly work and reliance on the voluntary reporting of infected cases (Park, Choi, and Ko 2021).

- *Partially centralized apps*: France ("StopCovid/ TousAntiCovid"), Singapore ("TraceTogether"), Australia ("COVIDSafe"), and New Zealand ("NZ COVID Tracer") released apps in which encounter logging is decentralized but infection reporting is centralized to support public health authorities. Iceland ("Rakning C-19") and Utah ("Healthy Together") developed apps supporting GPS-based contact tracing and centralized infection reporting.

- *Centralized tracking system*: Israel issued emergency regulations allowing its security agency (Shin Bet) to track mobile phones. The High Court of Justice issued several rulings on this: (i) holding in its April 26, 2020 decision that, while the initial reliance on the emergency regulation was constitutional, the government should start legislating a statute by April 30 to continue tracking (HCJ/2109/20, HCJ/2135/20, HCJ 2141/20, *Shachar Ben Meir, Adv. et al. v. Prime Minister et al.*, April 26, 2020), and (ii) ordering, in its March 1, 2021 decision, Shin Bet to end the indiscriminate use of the tracking system by March 14 and limit it to confirmed cases refusing epidemiological investigation (HCJ/6732/20, *Association for Civil Rights in Israel et al. v. Knesset et al.*, March 1, 2021.). Korea diverted an existing smart city hub to the Epidemic Investigation Support System, which pulls confirmed cases' location data (cell tower IDs) from mobile carriers and card transaction data from credit card companies (Park, Choi and Ko 2021). China deploys centralized "health code" (*jiankangma*) apps developed by Alipay and Tencent, which collects medical data and travel and contact history and generates colored QR codes identifying health risks (NHC/NHSA/NATCM 2020).

*Transparency and Control*

Valuing resilience over liberty regarding policing, public health, and emergency response, a smart city could sometimes consider a system not decentralized and processing personal data, especially under a more utilitarian than liberal tradition. When the system poses surveillance risks, citizens would demand data governance, as a precondition for public support and trust. Its key would be overall transparency and individual control.

- *Transparency*: Transparent data flow is the lynchpin of citizen trust and control. To overcome the limitations of unilateral disclosure such as privacy policies (Ben-Shahar and Schneider 2011), smart cities could consider more proactive measures such as impact assessment, oversight, audit, and the disclosure of source codes. The GDPR requires data controllers to conduct a data protection impact assessment (DPIA) if data processing is likely to result in a high risk to individuals (art. 35) but does not require it to be published or audited. However, for COVID-19 apps, several European countries made the DPIA or source codes publicly available. In Waterfront Toronto, advisory panel members resigned citing a lack of transparency and surveillance concerns in October 2018 (Cecco 2019). In response, Sidewalk Labs formulated a plan to conduct a publicly auditable Responsible Data Use Assessment and have urban data controlled by an independent entity called the Urban Data Trust (Sidewalk Labs 2019a). In November 2019, Sidewalk Labs withdrew the plan, amid criticism that it lacked oversight and contravened Canada's privacy law (Sidewalk Labs 2019b), eventually leading to the cancellation of the entire project in 2020.

- *Control*: A citizen's ability to control data more substantially and dynamically than giving one-time consent, such as through dashboards for adjusting privacy settings (FTC 2015), can engender public trust. The right to portability, data subjects' right under the GDPR to receive their personal data in a "structured, commonly used and machine-readable format" or transmit it to a third party they designate (art. 20), also confers more proactive control than a consent scheme.

**Predictive Modelling: Due Process, Fairness and Transparency**

*Overview*

A smart city's algorithm trains a model based on a training set split from the sensed and profiled data. The model can be used for predicting citizen behaviours and urban environments if found sufficiently accurate on a test set. By feeding the sensed and profiled data into a geospatial model visualized through 3D rendering and augmented reality, the model is evolving into a *Digital Twin*, a virtual clone of a city that simulates urban environments, infrastructure, and actions (U.K. NIC 2017). The predictive capability empowers the smart city to rate, rank, and classify inhabitants. It enables predictive or proactive policing or sentencing based on a predicted likelihood to (re)commit crime (Verma and Dombrowsk 2018), smart healthcare based on disease risk prediction (Nasr et al. 2021), and smart energy management based on predicted energy consumption (Amasyali and El-Goharv 2018).

The most notable example is *data-driven policing*, also called predictive, preventive, or proactive policing with slightly different nuances. Data-driven policing strategies include (i) a *person-focused strategy*, which predicts a small group of offenders posing a higher recidivism risk and intensively monitors them, (ii) a *location-based strategy*, which predicts crime hot spots using geographic information systems and intensively controls the spots, and (iii) a *strategy based on aggregate data*, such as search terms or topic trends (LAPC 2019). Los Angeles Police Department implemented, starting from August 2018, the person- and location-focused Los Angeles Strategic Extraction and Restoration (LASER) program, the location-focused Predictive Policing (PredPol) program, and a trend-based survey program developed by Elucd (*Id.*). However, it stopped LASER in 2019 following the Los Angeles Police Commission's audit which found a lack of oversight and inconsistency (*Id.*) and PredPol in 2020 due to financial constraints during the pandemic. New York Police Department disclosed its use of predictive policing in 2018 and has been implementing the Domain Awareness System since 2008 (Levine et al. 2017). Since 2013, Chicago Police Department used the Strategic Subject List, which predicts potential pistol assailants and victims (Sanders, Hunt, and Hollywood 2016), but scrapped it in 2020 amid controversy about the fitness of the data used.

*Due Process*
Predictive modelling poses not only the privacy risks noted above but also a risk of due process violation.

- *Person-focused modeling*: A smart city may try to predict a citizen's behavior

based on their data profiled to treat them disadvantageously. The draft EU AI Regulation bans AI systems from providing social scoring of natural persons for general purposes by public authorities or on their behalf (art. 5(1)(c)). An administrative action based on person-focused modeling can violate substantive due process if it infringes their constitutional right, right to participate in the political process, or right as discrete and insular minorities (*United States v. Carolene Products Co.*, 304 U.S. 144 (1938)) by use of flawed models, including those learned from inaccurate, unrepresentative, or biased data.

- *Cohort modeling*: Location-based modeling enables the police to reinforce on-site patrol, install more streetlamps and cameras, and otherwise improve pedestrian environments, which would generally be less infringing than individual monitoring based on person-focused modeling. The Federated Learning of Cohorts (FLoC) assigns individuals to several cohorts at the client level and allows a server to access only cohort profiles and train a model based on them (Xiao and Karlin 2022); the FLoC not only provides privacy guarantees but can empower smart cities to implement less intrusive cohort modeling in broader areas which had previously not been clearly segmented.

- *Aggregate modeling*: A city's use of a model learned from aggregate data such as search terms or topic trends would entail a relatively low legal risk.

### *Fairness and Transparency*

Moreover, citizens would call for fairness in predictions to guarantee equality before the law. As antidiscrimination laws in major jurisdictions tend to prohibit not only disparate treatment but also disparate impact, smart cities must adopt not input-centric but *output-*

*centric approaches* (Barocas, Hardt, and Narayanan 2019). Specifically, simply withholding sensitive attributes (SAs; such as race or gender) from a feature vector ($X$) is insufficient, because some other features can still be correlated with the SA (the problem of proxies) (*Id.*). For example, a location-based model can still be racially biased in a highly segregated region, even if race is removed from the feature vector before training it. Fairness metrics have been proposed to make fairness measurable (Table 1) (*Id.*).

|  | Condition | Interpretation |
|---|---|---|
| Independence | If the prediction ($\hat{Y}$) is independent from SA ($\hat{Y} \perp A$). <br> i.e., $P(\hat{Y} = +\|A = a_0) = P(\hat{Y} = +\|A = a_1)$ for $\forall a_0, a_1 \in A$. | Equality in outcome |
| Separation | If the prediction is independent from SA ($A$) under the condition of a given ground truth ($y$) ($\hat{Y} \perp A\|Y$). <br> i.e., $P(\hat{Y} = +\|Y = y, A = a_0) = P(\hat{Y} = +\|Y = y, A = a_1)$ for $\forall a_0, a_1 \in A, y \in \{+, -\}$. | Equality in opportunity: treating those having the same (de)merit equally regardless of SA |
| Sufficiency | If the ground truth ($Y$) is independent from SA under the condition of a given prediction ($\hat{y}$) ($Y \perp A\|\hat{Y}$). <br> i.e., $P(Y = +\|\hat{Y} = \hat{y}, A = a_0) = P(Y = +\|\hat{Y} = \hat{y}, A = a_1)$ for $\forall a_0, a_1 \in A, \hat{y} \in \{+, -\}$. | Calibration: those treated equally actually have the same (de)merit regardless of SA |

Table 1. Fairness Metrics (*Id.*)

More debates are needed for non-discrimination laws to convert current nebulous standards into more measurable and foreseeable rules embracing these metrics.

To satisfy citizens' right to know, a smart city must ensure not algorithmic transparency but model interpretability, as citizens' interest is not in the algorithm but the parameters constituting a model (how their characteristics or actions are correlated with the model's prediction). Various explainable AI (XAI) methodologies have been proposed to ensure model interpretability (Molnar 2022). However, counterfactual explanation appears the most promising instrument for fulfilling a legal duty to explain, as this can guide citizens regarding how much to change their characteristics or actions to be classified favourably (Wachter, Mittelstadt, and Russel 2017).

**Technological Determinism: The Problem of Human Autonomy**

Critics have accused smart cities of being trapped in "technological determinism" undermining human autonomy, reminiscent of Aldous Huxley's *Brave New World* (1932) and called for citizens' right to "unplugging" or self-determination (Calzada and Cobo 2015). Masdar and Songdo, earlier smart cities built under top-down schemes, were denounced as "stupefying" smart cities, which deaden the "people who live in its all-efficient embrace" (Sennett 2012). Urban technology is also suspected of exacerbating digital divide (Calzada and Cobo 2015). While these prospects are worth respectful attention, real-world smart cities have so far experienced the reverse – a low penetration or participation rate as seen in COVID-19 app cases. Cities have used gamification techniques to engage citizens in sustainable mobility solutions (Kzhamiakin et al. 2015) or garbage recycling (Briones et al. 2018) but also have realized the importance of actively reaching communities and residents. Implementing the Smart Columbus project funded by the U.S. Department of Transportation's Smart City Challenge, Columbus, Ohio shifted its focus from technology to residents to overcome hurdles including the low penetration of Pivot, a trip-planning app, and the interruption of self-driving Linden LEAP shuttles following an accident (City of Columbus 2021; McLean 2021).

A more material conflict between control and autonomy might concern smart mobility. The advent of mobility as a service (MaaS) will blur the distinction between individual car ownership and public transport. Cities are further planning to build vehicle-to-everything (V2X) networks, ITSs, and car and parking lot sharing applications, with which to connect, coordinate, and control self-driving cars, urban air mobility, and personal mobility, to manage traffic, save resources, and improve energy efficiency

(Xiong et al. 2012). Individual choices over mobility would be weakened, with the concept of autonomous driving shifting to controlled driving. Cities could restrict individual autonomy, such as manual driving or parking, since it causes each citizen's dominant strategy to deviate from a socially optimal route in pursuit of opportunistic payoffs, such as timesaving (Kearns and Roth 2019, pp. 105–115).

A smart city's shift toward distributed energy resources and prosumer participation is generally expected to foster energy autonomy (Rae and Bradley 2012). However, to maximize urban resilience, cities might have to restrain an individual preference for a mixture of centralized energy systems, which citizens might reveal particularly when grid parity and price reductions in energy storage systems have not yet been attained (McKenna 2018).

**Deprivatizing Digital Infrastructure: Crowding Out the Private Sector and Distorting Markets?**

Smart cities are expanding the concept of urban infrastructure to digital infrastructure with different layers, such as networks, platforms, and applications. IBM, Cisco, and other system integrators have played a pivotal role in developing and operating digital infrastructure for smart cities. Digital platforms have also penetrated smart city projects, as illustrated by Sidewalk Labs, Microsoft CityNext, and Alibaba City Brain. However, municipalities and central governments have also attempted to self-manage digital infrastructure. Several motivations drive this. A key rationale for municipal broadband, internet access provided by municipalities, has been to bridge the digital divide. The Federal Communications Commission (FCC), in its 2000 Second Broadband Progress Report, recognized direct public investment as one of the best practices for increasing access (FCC 2000). When NYC launched LinkNYC, free public Wi-Fi, in 2016, the city emphasized it would increase equity by increasing Wi-Fi access by low-income citizens, particularly Black and Latino residents (NYC 2016). Economic justifications are sometimes given. Korea, for its U-City initiatives, had outsourced its nationwide smart city hub to multinational companies, but in 2008, it shifted to self-development to "substitute import, reduce development costs, and ensure interoperability and compatibility between municipality systems" (MOLIT 2020).

Nevertheless, particularly municipal broadband has received criticism regarding crowding out the private sector and distorting competition. A transatlantic divide exists concerning how legal controversies develop.

- *Conflict with the state power*: As of December 1, 2021, 18 U.S. states restrict municipal broadband, and five additional states make it more difficult for municipalities to provide broadband (Cooper 2021), based on the U.S. Supreme Court's 2014 decision that not the FCC but each state has the authority to legalize or ban municipal broadband (*Nixon v. Missouri Municipal League*, 541 U.S. 125 (2004)). Seoul launched free public Wi-Fi named *Kkachi On* in 2020 but stopped it in 2022, transferring businesses to carriers, pursuant to a central telecommunications agency's ruling outlawing municipal broadband (Lim 2022).

- *Conflict with regional treaties blocking state aids*: The Treaty on the Functioning of the European Union (TFEU) deems any aid granted by a member state or through state resources, which can distort competition, incompatible with the European single market (art. 107(1)). Controversy has arisen regarding whether municipal broadband violates the state aid rules. The EC's 2013 Broadband Guidelines recognized compatibility on certain conditions including common interest, market failure, incentive effect, and transparency. On December 12, 2022, the EC adopted a new guideline allowing member states to invest in areas "where the market does not and is not likely to provide end users with a download speed of at least 1 Gbps and an upload speed of at least 150 Mbps" (Guidelines on State aid for broadband networks (2023/C 36/01), para. 54).

Some cities are expanding to the application layer. Rio de Janeiro developed a taxi-hailing app named Taxi.Rio, which, as of the end of 2021, has attracted 30 cities' interest, including four cities already using it (Diário do Rio 2021). As of the end of 2021, Korean municipalities have been directly managing at least 20 food delivery apps, two taxi-hailing apps, and an accommodation booking app, most of which have nominal active users (Bae 2021). Public intervention sometimes involves data pooling. In February 2017, NYC's Taxi & Limousine Commission approved regulations requiring ride-hailing companies to share journey data with the Commission (§§78-17(b), 78-21(e), TLC Rules and Local Laws: Chapter 78).

Reviewing the history of the privatization of telecommunication, discreet approaches are sensible in rewinding it. Less costly and less market-disturbing policy tools than deprivatization are available, such as universal service funds (Lyons 2018). The causes of market failure in the telecommunications industry, such as single-firm dominance, are not apparent in the digital platform industry (Hovenkamp 2021). Before cities directly compete with the private sector spending taxpayers' money, they must prioritize less disruptive tools such as antitrust enforcement.

**Outlook**

The discussions so far show that throwing technology at cities, albeit sometimes helpful, cannot always address concerns regarding civil liberties. Evidently, smart cities must pursue a Pareto improvement transcending trade-offs between resilience and liberties, by deploying privacy-enhancing technologies, cohort or aggregate modelling, fair machine learning, XAI, and other technical breakthroughs. Should that not work, however, cities must measure and compare costs and benefits involving resilience and liberties and seek a political consensus about the way they are balanced. A key to winning public trust amid intrinsic incommensurability could be to ensure transparency and control and to foster citizen participation in urban governance. European smart cities' experiments on living labs, innovative systems with living social settings based on citizen feedback loops, are establishing new possibilities for urban governance (Dutilleul, Birrer and Mensink 2010). The bottom-up approach is often a daunting task and might look inefficient and time-consuming, but it would make cities genuinely smart and eligible for our descendants' habitats.

## Disclosure Statement

This work was funded by the 2023 Research Fund (Joint Research) of the Seoul National University Law Research Institute, donated by the Seoul National University Law Foundation, and the National Research Foundation grant (No. 2022R1A5A708390811, Trustworthy AI).## References

K. Amasyali and N.M. El-Gohary, "A Review of Data-driven Building Energy Consumption Prediction Studies," *Renewable and Sustainable Energy Reviews* 81 (2018)1192–1205.

Y. An, S. Lee, S. Jung, H. Park, Y. Song, and T. Ko, "Privacy-Oriented Technique for COVID-19 Contact Tracing (PROTECT) Using Homomorphic Encryption: Design and Development Study," *Journal of Medical Internet Research* 23:7 (2021) e26371.

Apple/Google, "Exposure Notifications: Help Slow the Spread of COVID-19, with One Step on Your Phone." <https://www.google.com/covid19/exposurenotifications/> Accessed March 25, 2023.

C. Arnett, "From Decarceration to E-carceration," *Cardozo Law Review* 41:2 (2019) 641–720.

Article 29 Data Protection Working Party, Opinion 05/2014 on Anonymisation Techniques (0829/14/EN WP216) (2014).

H.J. Asghar, P. Tyler, and M.A. Kaafar, "Differentially Private Release of Public Transport Data: The Opal Use Case," *arXiv* (2017). <https://arxiv.org/pdf/1705.05957.pdf>

D. Bae, "The Average Daily User of 20 Municipal Delivery Apps, which Absorbed KRW Billions of Budget Spending, is only 3% of that of BAEMIN App," *Chosun Biz,* October 7, 2021.

<https://biz.chosun.com/distribution/channel/2021/10/07/DV3SEUJRHNCITOIXZQMQSZKJHY/> Accessed March 25, 2023.

S. Barocas, M. Hardt, and A. Narayanan, *Fairness and Machine Learning: Limitations and Opportunities* (2019). <https://fairmlbook.org> Accessed March 25, 2023.

R. Bassily and A. Smith, "Local, Private, Efficient Protocols for Succinct Histograms," *Annual ACM Symposium on Theory of Computing (STOC '15)* (2015) 127–135.

O. Ben-Shahar, "Data Pollution," *Journal of Legal Analysis* 11 (2019), 104–159.

O. Ben-Shahar and C.E. Schneider, "The Failure of Mandated Disclosure," *University Pennsylvania Law Review* 159 (2011) 647–749.

A.G. Briones, P. Chamoso, A. Rivas, S. Rodríguez, F.D.L. Prieta, J. Prieto, and J.M. Corchado, "Use of Gamification Techniques to Encourage Garbage Recycling: A smart City Approach," *Knowledge Management in Organisations (KMO '18)* (2018) 674–685.

C. Buchanan, *Traffic in Towns: A Study of the Long Term Problems of Traffic in Urban Areas*, (London: Her Majesty's Stationery Office, 1963).

Busan Eco Delta City (EDC), "Call for Smart Village Resident Application," Nov 11, 2020. <https://busan-smartvillage.com/apply> Accessed March 25, 2023.

I. Calzada and C. Cobo, "Unplugging: Deconstructing the Smart City," *Journal of Urban Technology* 22:1 (2015), 23–43.

C. Catlett, P. Backman, H. Nusbaum, M.E. Papka, M.G. Berman, and R. Sankaran, "Measuring Cities with Software-Defined Sensors," *Journal of Social Computing* 1:1 (2020) 14–27.

A. Cavoukian, "Privacy by Design: The 7 Foundational Principles" (2011). <https://www.ipc.on.ca/wp-content/uploads/Resources/7foundationalprinciples.pdf> Accessed March 25, 2023.

L. Cecco, "'Surveillance Capitalism': Critic Urges Toronto to Abandon Smart City Projects," *The Guardian*, June 6, 2019.

R. Chandran, "How Japan's 'Opt-in' Smart City Could Change Urban Living," *Global Technology Governance Summit 2021*, March 15, 2021. <https://www.weforum.org/agenda/2021/03/japanese-smart-city-residents-privacy-protection-data/> Accessed March 25, 2023.

T. Chiba, R. Inoue, and J. Miura, "Toyota Begins Building its Own Prototype Smart City 'of the Future'", *The Asahi Simbum*, Feb 24, 2021.

City and County of San Francisco, "DataSF: Resources – Open Data Release Toolkit: Privacy Edition (Ver 1.2)" (2016). <https://datasf.org/resources/open-data-release-toolkit/> Accessed March 25, 2023.

City of Columbus, "Final Report for the Smart Columbus Demonstration Program," June 15, 2021. <https://smart.columbus.gov/programs/smart-city-demonstration> Accessed March 25, 2023.

T. Cooper, "Municipal Broadband is Restricted in 18 States across the U.S. in 2021," *Broadbandnow Research*, Dec 1, 2021. <https://broadbandnow.com/report/municipal-broadband-roadblocks/> Accessed March 25, 2023.

J. Corburn, "Confronting the Challenges in Reconnecting Urban Planning and Public Health," *American Journal of Public Health* 94 (2004) 541–546.

Datatilsynet, *Varsel om vedtak om midlertidig forbud mot å behandle personopplysninger - appen Smittestopp*, June 12, 2020. <https://www.datatilsynet.no/aktuelt/aktuelle-nyheter-2020/midlertidig-stans-av-appen-smittestopp/> Accessed March 25, 2023.

Diário do Rio, *30 cidades já mostraram interesse pelo Taxi.Rio*, Dec 30, 2021. <https://diariodorio.com/30-cidades-ja-mostraram-interesse-pelo-taxi-rio/> Accessed March 25, 2023.

B. Dutilleul, F.A. Birrer, and W. Mensink, "Unpacking European Living Labs: Analysing Innovation's Social Dimensions," *Central European Journal of Public Policy* 4:1 (2010) 60–85.

C. Dwork and A. Roth, "The Algorithmic Foundations of Differential Privacy," *Foundations and Trends in Theoretical Computer Science* 9:3–4 (2013) 211–407.

C. Dwork, "Differential Privacy: A Survey of Results," *Theory and Applications of Models of Computation. (TAMC '08)* (2008) 1–19.

G. Eibl and D. Engel, "Differential Privacy for Real Smart Metering Data," *Computer Science – Research and Development* 32 (2017) 173–182.

J. Eberhardt and S. Tai, "On or Off the Blockchain? Insights on Off-chaining Computation and Data," *European Conference on Service-Oriented and Cloud Computing (ESOCC '17)* (2017) 3–15.

Ú. Erlingsson, V. Pihur, and A. Korolova, "RAPPOR: Randomized Aggregatable Privacy-preserving Ordinal Response," *ACM SIGSAC Conference on Computer and Communications Security (CCS '14)* (2014) 1054–1067.

European Data Protection Board (EDPB), Guidelines 04/2020 on the use of location data and contact tracing tools in the context of the COVID-19 outbreak, April 21, 2020.

Federal Communications Commission (FCC), *Deployment of Advanced Telecommunications Capability: Second Report*, August 2000.

Federal Trade Commission (FTC), "The Internet of Things: Privacy and Security in a Connected World" (2015).

A. Garcia-Garcia, S. Orts-Escolano, S. Oprea, V. Villena-Martinez, and J. Garcia-Rodriguez, "A Review on Deep Learning Techniques Applied to Semantic Segmentation," *arXiv* (2017). <https://arxiv.org/abs/1704.06857>

C. Gentry, "Computing Arbitrary Functions of Encrypted Data," *Communications of the ACM* 53:3 (2010) 97–105.

R.C. Geyer, T. Klein, and M. Nabi, "Differentially Private Federated Learning: A Client Level Perspective," *arXiv* (2017). https://arxiv.org/abs/1712.07557?context=cs.LG>

E.L. Glaeser and B. Sacerdote, "Why is There More Crime in Cities?" *Journal of Political Economy* 107.S6 (1999) S225–S258.

Google, "Tackling Urban Mobility with Technology," *Google Europe Blog*, 2015. <https://europe.googleblog.com/2015/11/tackling-urban-mobility-with-technology.html> Accessed March 25, 2023.

M.U. Hassan, M.H. Rehmani, R. Kotagiri, J. Zhang, and J. Chen, "Differential Privacy for Renewable Energy Resources Based Smart Metering," *Journal of Parallel and Distributed Computing* 131 (2019) 69–80.

A. Henneguelle, B. Monnery, and A. Kensey, "Better at Home than in Prison? The Effects of Electronic Monitoring on Recidivism in France," *Journal of Law & Economics* 59:3 (2016), 629–667.

H. Hovenkamp, "Antitrust and Platform Monopoly," *Yale Law Journal* 130:8 (2021) 1952–2273.

J. Kaye, L.B. Moraia, L. Curren, J. Bell, C. Mitchell, S. Soini, N. Hoppe, M. Øien, and E. Rial-Sebbag, "Consent for Biobanking: The Legal Framework of Countries in the BioSHaRE-EU Project," *Biopreservation and Biobanking* 14:3 (2016), 195–200.

M. Kearns and A. Roth, *The Ethical Algorithm: The Science of Socially Aware Algorithm Design* (New York: Oxford University Press, 2019).

J. Konecny, H. McMahan, D. Ramage, and P. Richtarik, "Federated Optimization: Distributed Machine Learning for On-device Intelligence," *arXiv* (2016). <https://arxiv.org/abs/1610.02527>

R. Kzhamiakin, A. Marconi, M. Perillo, M. Pistore, G. Valetto, L. Piras, F. Avesani, and N. Perri, "Using Gamification to Incentivize Sustainable Urban Mobility," *IEEE International Smart Cities Conference (ISC2 '15)* (2015).

Los Angeles Police Commission (LAPC), Office of the Inspector General, *Review of Selected Los Angeles Police Department Data-Driven Policing Strategies*, 2019. <http://www.lapdpolicecom.lacity.org/031219/BPC_19-0072.pdf> Accessed March 25, 2023.

N.T. Lee and C. Chin, "Police Surveillance and Facial Recognition: Why Data Privacy is Imperative for Communities of Color," *Brookings Institution* (2022). <https://www.brookings.edu/research/police-surveillance-and-facial-recognition-why-data-privacy-is-an-imperative-for-communities-of-color/> Accessed March 25, 2023.

E.S. Levine, J. Tisch, A. Tasso, and M. Joy, "The New York City Police Department's Domain Awareness System," *INFORMS Journal on. Applied Analytics* 47:1 (2017) 70–84.

F. Li, B. Luo, and P. Liu, "Secure Information Aggregation for Smart Grids Using Homomorphic Encryption," *IEEE International Conference on Smart Grid Communications (SmartGridComm)* (2010) 327–332.

N. Li, T. Li, and S. Venkatasubramanian, "$t$-Closeness: Privacy beyond $k$-Anonymity and $l$-Diversity," *International Conference on Data Engineering (ICDE '07)* (2007) 106–115.

H. Lim, "Seoul Scraps *Kkachi On* Business … Cooperating with the Ministry of Science and ICT," *The Asia Business Daily*, Feb 24, 2022. <https://www.asiae.co.kr/article/2022022312313003718> Accessed March 25, 2023.

D.A. Lyons, "Narrowing the Digital Divide: A Better Broadband Universal Service Program," *U.C. Davis Law Review* (2018) 52, 803–854.

A. Machanavajjhala, J. Gehrke, D. Kifer, and M. Venkitasubramaniam, "$\ell$-Diversity: Privacy beyond $k$-Anonymity," *International Conference on Data Engineering (ICDE '06)* (2006).

A. Machanavajjhala, D. Kifer, J. Abowd, J. Gehrke, and L. Vilhuber, "Privacy: Theory Meets Practice on the Map," *International Conference on Data Engineering (ICDE '08)* (2008) 277–286.

R. McKenna, "The Double-edged Sword of Decentralized Energy Autonomy," *Energy Policy* 113 (2018) 747–750.

D. McLean, "6 Questions with Smart Columbus on Lessons since Program's Conclusion," *SmartcitiesDive*, Oct 11, 2021. <https://www.smartcitiesdive.com/news/smart-columbus-ohio-smart-city-challenge-winner-lessons-learned/607999/> Accessed March 25, 2023.

R. McPherson, R. Shokri, and V. Shmatikov, "Defeating Image Obfuscation with Deep Learning," *arXiv*, 2016. <https://doi.org/10.48550/arXiv.1609.00408>.

Ministry of Land, Infrastructure, and Transport, Korea (MOLIT), *Laying the Basis for the Smart City Integrated Platform* (2020). <https://smartcity.go.kr/2020/06/11/%ec%8a%a4%eb%a7%88%ed%8a%b8%ec%8b%9c%ed%8b%b0-%ed%86%b5%ed%95%a9%ed%94%8c%eb%9e%ab%ed%8f%bc-%ea%b8%b0%eb%b0%98%ea%b5%ac%ec%b6%95/> Accessed March 25, 2023.

C. Molnar, *Interpretable Machine Learning: A Guide for Making Black Box Models Explainable* (2022). <https://christophm.github.io/interpretable-ml-book/> Accessed March 25, 2023.

M. Naphade, G. Banavar, C. Harrison, J. Paraszczak, and R. Morris, "Smarter Cities and Their Innovation Challenges," *Computer* 44:6 (2011) 32–39.

M. Nasr, M. Islam, S. Shehata, F. Karray, and Y. Quintana, "Smart Healthcare in the Age of AI: Recent Advances, Challenges, and Future Prospects," *IEEE Access* 9 (2021) 145248–145270.

National Health Commission, National Healthcare Security Administration, and National Administration of Traditional Chinese Medicine (NHC/NHSA/NATCM), *Notice on In-depth Implementation of "Internet + Medical Health" and "Five Ones" Service Actions*, Dec 4, 2020.

<http://www.gov.cn/zhengce/zhengceku/2020-12/10/content_5568777.htm> Accessed March 25, 2023.

New York City (NYC), "Mayor de Blasio Announces Public Launch of LinkNYC Program, Largest and Fastest Free Municipal Wi-Fi Network in the World," Feb 18, 2016. <https://www1.nyc.gov/office-of-the-mayor/news/184-16/mayor-de-blasio-public-launch-linknyc-program-largest-fastest-free-municipal#/0> Accessed March 25, 2023.

New York City (NYC), "Guidelines for the Internet of Things" (2017). <https://iot.cityofnewyork.us/privacy-and-transparency/> Accessed March 25, 2023.

Organization for Economic Co-operation and Development (OECD), *Divided Cities: Understanding Intra-urban Inequalities* (Paris: OECD Publishing, 2018).

S. Park, G.J. Choi, and H. Ko, "Privacy in the Time of COVID-19: Divergent Paths for Contact Tracing and Route-Disclosure Mechanisms in South Korea," *IEEE Security & Privacy* 19:3 (2021) 51–56.

T.E. Raghunathan, "Synthetic Data," *Annual Review of Statistics and its Application* 8 (2021) 129–140.

A.G. Roy, S. Siddiqui, S. Poelsterl, N. Navab, and C. Wachinger, "BrainTorrent: A Peer-to-peer Environment for Decentralized Federated Learning," *arXiv* (2019), <https://arxiv.org/pdf/1905.06731.pdf>

J. Sanders, P. Hunt, and J.S. Hollywood, "Predictions Put into Practice: A Quasi-experimental Evaluation of Chicago's Predictive Policing Pilot," *Journal of Experimental Criminology* 12 (2016) 347–371.

J. Sadowski and F.A. Pasquale, "The Spectrum of Control: A Social Theory of the Smart City," *First Monday* 20:7 (2015).

R. Sennett, "No One Likes a City That's Too Smart," *The Guardian*, Dec 4, 2012.

Sidewalk Labs, "Master Innovation and Development Plan: Toronto Tomorrow – A New Approach for Inclusive Growth" (2019a). <https://www.sidewalklabs.com/toronto> Accessed March 25, 2023.

Sidewalk Labs, "Master Innovation and Development Plan: Digital Innovation Appendix" (2019b). <https://www.sidewalklabs.com/toronto> Accessed March 25, 2023.

M. Sookhak, H. Tang, Y. He and F.R., Yu, "Security and Privacy of Smart Cities: A Survey, Research Issues and Challenges," *IEEE Communications Surveys & Tutorials* 21:2 (2019) 1718–1743.

Office of the Privacy Commissioner of Canada, "Consent and Privacy: A Discussion Paper Exploring Potential Enhancements to Consent under the Personal Information Protection and Electronic Document Act" (2016).

C. Rae and F. Bradley, "Energy Autonomy in Sustainable Communities – A Review of Key Issues," *Renewable and Sustainable Energy Reviews* 16:9 (2012) 6497–6506.

Reuters, "U.S. Cities are Backing off Banning Facial Recognition as Crime Rises," May 13, 2022.

P. Samarati and L. Sweeney, "Protecting Privacy When Disclosing Information: *k*-Anonymity and its Enforcement through Generalization and Suppression" (1998). <https://dataprivacylab.org/dataprivacy/projects/kanonymity/paper3.pdf> Accessed March 25, 2023.

W. Shi, J. Cao, Q. Zhang, Y. Li, and L. Xu, "Edge Computing: Vision and Challenges," *IEEE IoT Journal* 3:5 (2016) 637–646.

Smart Nation Singapore, "Digital Contract Tracing" (2022). <https://www.smartnation.gov.sg/combating-covid-19/digital-contact-tracing> Accessed March 25, 2023.

State of Florida Office of the Auditor General, "Performance Audit of the Use of Electronic

Monitoring within the Community Control Program Administered by the Department of Corrections," Feb 17, 1993. <https://www.ojp.gov/ncjrs/virtual-library/abstracts/performance-audit-use-electronic-monitoring-within-community> Accessed March 25, 2023.

United Kingdom National Infrastructure Commission (U.K. NIC), "Data for the Public Good" (2017) 60–66. <https://nic.org.uk/app/uploads/Data-for-the-Public-Good-NIC-Report.pdf> Accessed March 25, 2023.

United Nations (UN), Department of Economic and Social Affairs, *World Urbanization Prospects: The 2018 Revision* (New York: United Nations, 2019).

United Nations (UN), Human Settlements Programme, *World Cities Report 2020: The Value of Sustainable Urbanization* (New York: United Nations, 2020).

N. Verma and L. Dombrowsk, "Confronting Social Criticisms: Challenges When Adopting Data-driven Policing Strategies," *Conference on Human Factors in Computing Systems (CHI '18)* (2018) 1–13.

S. Wachter, B. Mittelstadt, and C. Russell, "Counterfactual Explanations without Opening the Black Box: Automated Decisions and the GDPR," *Harvard Journal of Law and Technology* 31:2 (2017) 841–887.

T. Wang, J. Blocki, N. Li, and S. Jha, "Locally Differentially Private Protocols for Frequency Estimation," *USENIX Security Symposium (SEC '17)* (2017) 729–745.

D. Warren II and L. Brandeis, "The Right to Privacy," *Harvard Law Review* 4:5 (1890) 193–220.

Y. Xiao and J. Karlin, "Federated Learning of Cohorts" (2022). <https://wicg.github.io/floc/> Accessed March 25, 2023.

Z. Xiong, H. Sheng, W. Rong, and D.E. Cooper, "Intelligent Transportation Systems for Smart Cities: A Progress Review," *Science China Information Sciences* 55 (2012) 2908–2914.

Y. Yang, X. Huang, X. Liu, H. Cheng, J. Weng, X. Luo, and V. Chang, "A Comprehensive Survey on Secure Outsourced Computation and its Applications," *IEEE Access* 7 (2019) 159426–159465.

A.C. Yao, "How to Generate and Exchange Secrets," *IEEE Annual Symposium on Foundations of Computer Science (FOCS)* (1986) 162–167.

A.U. Zaman and S. Lehmann, "The Zero Waste Index: A Performance Measurement Tool for Waste Management Systems in a 'Zero Waste City'," *Journal of Cleaner Production* 50 (2013) 123–132.

Z. Zheng, Y. Zhou, Y. Sun, Z. Wang, B. Liu, and K. Li, "Applications of Federated Learning in Smart Cities: Recent Advances, Taxonomy, and Open Challenges," *Connection Science* 34:1 (2022) 1–28.